\documentclass[reprint, nofootinbib]{revtex4-2}

\usepackage{dcolumn}
\usepackage{bm}
\usepackage{indentfirst}
\usepackage{mathrsfs}
\usepackage{comment}
\usepackage{amsfonts}
\usepackage{amssymb}
\usepackage{dsfont}
\usepackage{euscript}
\usepackage{graphicx}
\usepackage{wrapfig}
\usepackage{mathtools}
\usepackage{pifont}
\usepackage[dvipsnames]{xcolor}
\usepackage{tensor}
\usepackage{amsmath,latexsym,esint}
\usepackage{cmap}
\usepackage[caption=false]{subfig}
\usepackage{hyperref}
\usepackage{float}
\usepackage{xcolor}
\usepackage[toc,page]{appendix}

\begin{document}
	\preprint{}
	\title{Taub-NUT as gravitational dyon with torsion.
     }
    
	\author{Dmitri Gal'tsov}
	\email{galtsov@phys.msu.ru}
	\author{Rostom Karsanov}
	\email{karsanovrz@my.msu.ru}
	\affiliation{Faculty of Physics, Moscow State University, 119899, Moscow, Russia
	}

\begin{abstract} 
We show that the Levi-Civita connection for the Taub-NUT metric is singular on Misner strings and generates an Einstein tensor containing the squares of the delta function. This can be improved by introducing torsion, in which case the Einstein tensor may be interpreted as that of a cosmic string with negative variable tension. Komar flows in Misner strings are generated by torsion, which is therefore responsible for the asymptotic Komar charges. The Taub-NUT metric then appears as a soliton supported by two spin-fluid beams.

\end{abstract}
	
\maketitle

\section{Introduction}
Although more than half a century has passed since the discovery of the Taub-NUT metric \cite{Newman:1963yy} and the significance of Misner strings (MS) \cite{Misner:1963fr}, there is still no consensus on the nature of these objects. Misner \cite{Misner:1963fr} proposed eliminating strings by imposing a periodic condition on time, but this only exacerbated the problem \cite{Miller:1971em}. Bonnor \cite{Bonnor:1969ala,Sackfield71} proposed considering the MS as weak singularities similar to cosmic strings, and Dowker \cite{Dowker:1974znr} declared the Taub-NUT solution to be a gravitational dyon. But the question of what matter can give rise to Misner strings in Bonnor's interpretation remains a subject of debate. Calculation of the outgoing Komar fluxes across the cylinders surrounding the strings \cite{Manko:2005nm} seemed to indicate the presence of negative mass rods with infinite angular momenta as the source of the MS, but a more detailed analysis \cite{Galtsov:2026wxl}, based on the concept of Komar gravitational strength tensor \cite{Simon:1984qb}, showed that the flux of the corresponding gravitational field lines originates from the positive horizon mass but directed inside the cylinders surrounding MS, explaining the negative sign in \cite{Manko:2005nm}.

Bonnor's second proposal, put forward in 2001 \cite{Bonnor:2001mh,bonnor2001} and discussed in \cite{Griffiths:2009dfa}, was that MS are torsion singularities similar to those of rotating cosmic strings with dislocations \cite{Tod1994,Letelier:1995ze} proposed in \cite{Galtsov:1993ne}. Such singularities are typically surrounded by a chronology-violating region where the azimuthal angle $\varphi$ becomes a timelike coordinate \cite{Ozdemir:2005tb}. The torsion proposal received considerable support in the recent discussion of gravitational S-duality \cite{Henneaux:2004jw,Bunster:2006rt,Argurio:2008zt,Argurio:2009xr,Kol:2020zth}. It was noted that for a nonlinear gravitational S-duality to exist, torsion must be taken into account due to the possible violation of the algebraic Bianchi identity by the Taub-NUT solution \cite{Kol:2023yxd}, and the question of what the gravitational dyon is in Einstein-Cartan theory was raised. The nature of the Misner singularities in this light was also discussed in \cite{Kolar:2025kle}, where the observer-dependence of their conical structure was revealed.

In the previous work \cite{Galtsov:2026wxl}, we discovered that the Komar two-form  \cite{Simon:1984qb}, corresponding to the timelike Killing vector, develops distributional fluxes in the MS that can enter and exit through their lateral surfaces, explaining the discrepancy between the Komar fluxes across the horizon and at infinity. Here, we extend this to the azimuthal Killing vector, revealing balance of the angular momentum and offering new arguments in favor of Bonnor's second (torsional) proposal. Specifically, we develop a Cartan description that leads to a dilemma: either no torsion but a singular connection, leading to an Einstein tensor with delta-squared terms, or singular torsion with a regular connection, leading to an   Einstein tensor with only linear delta-function terms  and offering an explanation for Komar's singular flows.

 \section{Komar strength tensors}
The Taub-NUT metric is an exact solution to Einstein's vacuum equations with two parameters: the mass $m$ and the NUT parameter $n$:
\begin{equation}\label{TaubNUT}
\begin{aligned}
	&ds^2=-\frac{\Delta}{\Sigma}(dt+2n\cos \theta d\varphi)^2+\frac{\Sigma}{\Delta}dr^2+\Sigma d\Omega^2,\\
	&\Delta=r^2-2mr-n^2,\qquad \Sigma=r^2+n^2.
\end{aligned}
\end{equation}	
This metric has two commuting Killing vectors $k^\mu_{(a)}=\delta^\mu_a$, where $a=t,\varphi$, or, in operator notation, $k^\mu_{(a)}\partial_\mu=\left(\partial_t, \partial_\varphi\right)$. The first is timelike outside the horizon $r>r_H=m+\sqrt{m^2+n^2}$, and the second is spacelike outside the chronosphere boundary $r>r_c$, where $r_c$ is the root of the equation $g_{\varphi\varphi}=0$. When working with Killing objects in aggregate, we often omit the subscript $a$. 	
	The inverse metric has a singularity in the northern and southern MS, located at $\theta=0,\; r>r_H$ and $\theta=\pi,\; r>r_H$, respectively.
	
Killing vectors generate Komar charges inside the closed boundary $\partial\Sigma$ of some spatial region $\Sigma$:
\begin{equation}\label{Qkomar}
	Q_k[\Sigma]
	=
	\frac{1}{8\pi c}
	\oint_{\partial\Sigma}\star K,\qquad K =dk \;,
\end{equation}
where $\star$ denotes Hodge dualization, $ k=k^\mu g_{\mu\nu}dx^\nu$ -- the Killing one-forms. The mass $M$ corresponds to $\partial_t$ and $c=-1$, and the angular momentum $J$ corresponds to $\partial_\varphi$ and $c=2$. To maintain S-duality in the form of the exchange of mass and NUT $m\leftrightarrow n$, the   dual Komar charges has to be built too  by integrating $K$ itself  \cite{Bossard:2008sw}.   
	
Following \cite{Simon:1984qb,Galtsov:2026wxl} we treat the two-forms $K$ as field-strength tensors similar to the Maxwell tensor whose electric and magnetic fluxes give rise to the corresponding Komar charges. Traditionally, non-zero fluxes through cylindrical surfaces surrounding the MS have been interpreted as a sign of the presence of singular matter sources in the Misner-Bonnor strings, from which the corresponding K-field lines emanate. An alternative interpretation, proposed in \cite{Galtsov:2026wxl}, suggests that these field lines close inside the MS, forming loops without point sources.
Indeed, Killing one-forms can be represented as $k=fdt+gd\varphi$,
where $f,g$ are functions of $r,\theta$. As soon as $g\neq 0$ on the MS, $k$ becomes singular there, and its differential should be considered in the sense of distributions. Then the one-form $d\varphi$ is not closed, but 
\begin{equation}\label{ddphi}	
dd\varphi=2\pi\,\delta^2(\mathbf{x})\,dx\wedge dy,
\end{equation}
where $x,\;y$ are local Cartesian-like coordinates
$x=\sin \theta \cos \varphi,\;
y=\sin \theta \sin \varphi,
$ and $\delta^2(\mathbf{x})=\delta(x)\delta(y)$ is the two-dimensional delta-function. For a simple derivation of this formula see \cite{Felsager:1981iy} and
\cite{Galtsov:2026wxl}, for deeper mathematical grounds see \cite{derham1984differentiable}.
This relation implies that the Komar field $K=dk$ consists of a regular part $K^{\mathrm{reg}}$ and a singular part ${\cal K}^s$:
\begin{equation*}
\begin{aligned}	
	&K =K^{\mathrm{reg}} +{\cal K}^s,\quad K^{\mathrm{reg}}=df\wedge dt+dg\wedge d\varphi,\nonumber\\ &{\cal K}^s=2\pi g^{s} \delta^2(\mathbf{x})\,dx\wedge dy,
\end{aligned}
\end{equation*}
where the index $s=\pm$ denotes the values taken on the northern $(\cos\theta=1)$ and southern 
$(\cos\theta=-1)$ MS. Taking the second differential  with \eqref{ddphi} again, we obtain $d(K^{\mathrm{reg}} +{\cal K}^s)=0$ identically.  Here the differentiation of the regular part is performed using \eqref{ddphi}, while in the second singular term it is the regular coefficient part which is differentiated. This means that our force lines   are continuous, forming closed loops. Similar considerations hold for the one-form $\tilde{k}$, generating the Hodge-dual Komar form $\star K=d\tilde{k}$.
These are needed to maintain gravitational S-duality.
 
Let us consider the case of angular momentum in more detail, referring to \cite{Galtsov:2026wxl} for the case of mass.
The corresponding Killing one-form is:
\begin{equation}\label{1-form g}
	k_{(\varphi)}\!=\!-2\lambda\cos \theta (dt+2n\cos \theta d\varphi)+\Sigma \sin^2 \theta d\varphi,
\end{equation}
with $\lambda\!=\!n\Delta/\Sigma$.
For the regular part $K^{\mathrm reg}$, we obtain:
\begin{equation*}
    \begin{aligned}
	&K_{\theta t}=2\lambda\sin \theta,\;\; K_{\theta \varphi}=(\Sigma +4n\lambda)\sin 2\theta, \\ &K_{rt}=-4\lambda_n\cos \theta,\;\; K_{r \varphi}=2r\sin^2\theta -8n\lambda_n\cos^2\theta,
\end{aligned}
\end{equation*}
where we designated \cite{Galtsov:2026wxl}:
\begin{equation}\label{lamn}
\lambda_n=mn\frac{\tilde{\Delta}}{\Sigma^2},\quad   \tilde{\Delta}=r^2+2r\frac{n^2}{m}-n^2. 
\end{equation}  For the singular part, we obtain: 
\begin{equation}\label{sing L} \mathcal{K}^s_{(\varphi)}= - 8\pi n\lambda \delta^2(\mathbf{x})dx \wedge dy. 
\end{equation}
Similarly, introducing the dual
potential one-form \begin{equation}\label{1-form dual to g}
\tilde{k}_{(\varphi)}=-2r\cos \theta\frac{\Delta}{\Sigma}(dt+2n\cos \theta d\varphi)-n\Delta'\sin^2\theta d\varphi,
\end{equation}
we find the singular term: 
\begin{equation}\label{dual singular L}
	\tilde{{\cal K}}^s_{(\varphi)}= - 8\pi r \lambda \delta^2(\mathbf{x})dx \wedge dy.    
\end{equation}

\begin{figure}[H]
	\centering
\includegraphics[width=0.9\linewidth]{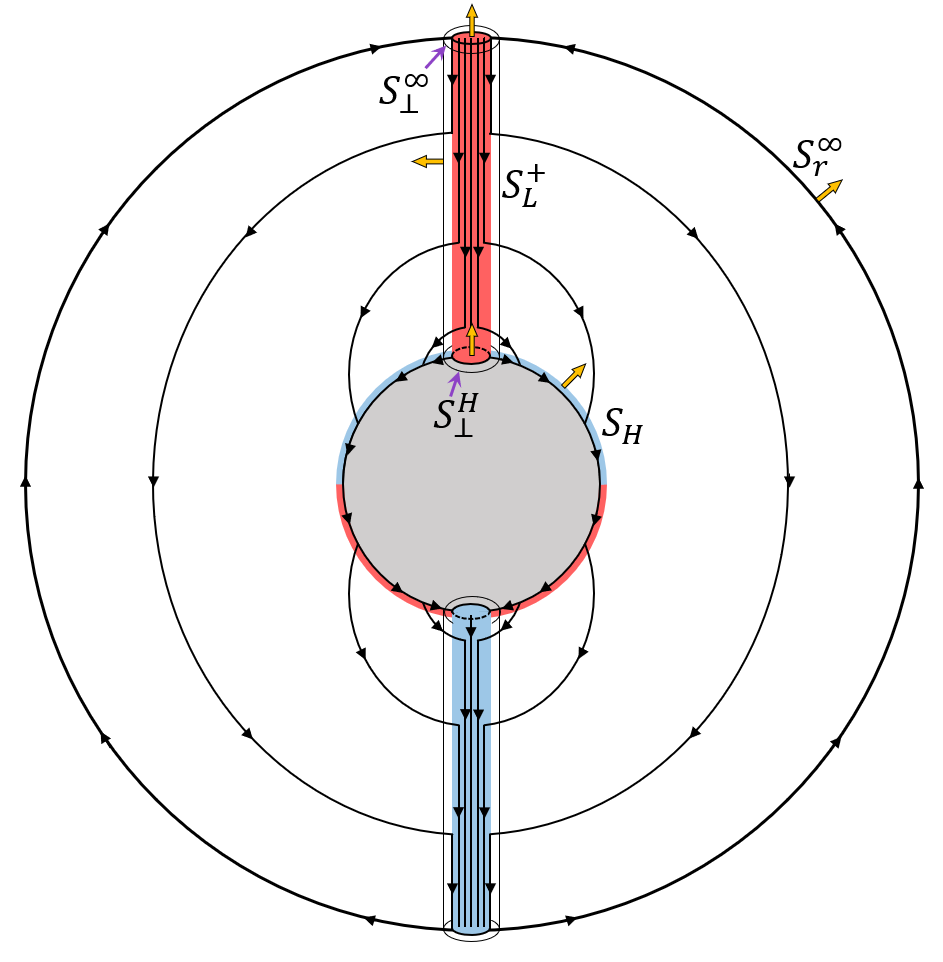}
	\caption{Schematic picture of the angular momentum Komar fluxes in Taub-NUT Misner strings closing in the bulk. Yellow arrows indicate the directions of the corresponding normal vectors to the surfaces of integration. Force lines escape from the northern Misner tubes (marked red) entering either into the southern string (marked blue) or the horizon. Other lines come from the part of horizon (red) then entering the southern string. The lines close on the horizon or at infinity (for more detailed explanations see \cite{Galtsov:2026wxl}) all forming closed loops. No momentum lines cross the infinite sphere, the total angular momentum being zero.}
	\label{FIG 1}
\end{figure}

Following \cite{Galtsov:2026wxl}, one can verify the balance equation for the field $\star K_{(\varphi)}$, say, for the northern string. The lateral $L$ and orthogonal $\perp$ fluxes are introduced in the MS cylinders, as shown in FIG.\ref{FIG 1}:
\begin{equation}
    \Phi_L=\int_{S_L}\star K_{(\varphi)},\qquad \Phi_\perp=\int_{S_\perp}\tilde{{\cal K}^+}.
\end{equation}
  These fluxes satisfy the relations:
\begin{equation}
    \Phi_L(r)=- \Phi_\perp(r)=8\pi r \lambda ,\quad\Phi_\perp(r_H)=0,
\end{equation}
so the total flux through the closed 2-surface is zero,  $\left[\Phi_\perp(r)\!-\!\Phi_\perp(r_H)\!+\!\Phi_L(r)\right]\!=\!16\pi J_+\!=\!0,
   $
and similarly for the southern string. 

The diverging angular momenta  in the Komar--Tomimatsu approach \cite{Manko:2005nm}  are caused by the force lines of   $\star K_{(\varphi)}$, escaping from the MS into the bulk  through the lateral surfaces only.
 To make contact with these  calculations we can  introduce  fictitious spin-densities by:  
\begin{equation}\label{J fict} 
	J^{\rm{fict}}_s=\frac{1}{16\pi}\Phi^s_L=\int^r_{r_H}\lambda_J^s dr,\;
\end{equation}
then one obtains the expression
\begin{equation}\label{Komar spin dens}
	\lambda_J^s=s\frac{n}{2}\left(1+\frac{2n^2\Delta}{\Sigma^2}\right). 
\end{equation}
At spatial infinity they tend to the constant values $\lambda_J^s=sn/2$, leading to divergent fictitious momenta \eqref{J fict}.

The flux of the bulk angular momentum field lines across the sphere of any radius vanishes
\begin{equation}
	\Phi_S=\int_{S_r}\star K_{(\varphi)} =0,  
\end{equation} 
so  fictitious angular momenta of the two MS cancel. But the horizon is ``polarized'':  
\begin{equation}
\!\!\!	J_H\!=\!\!\int_{S_H}\!\!\! \lambda_J^H\sqrt{\gamma_H}d\theta d\varphi,\quad
	\lambda_J^H\!=\!-\!\frac{n}{8\pi r_H}\cos \theta,
\end{equation}
while the integral over $\theta$ vanishes. Numerical pictires of angular momentum field lines are shown in FIG.\ref{FIG 2} including the rotation case (not treated in the main text):

\begin{figure}[H]
\centering
\subfloat[$m=0.1,\;n=2,\;a=0.$]{%
    \includegraphics[width=0.23\textwidth]{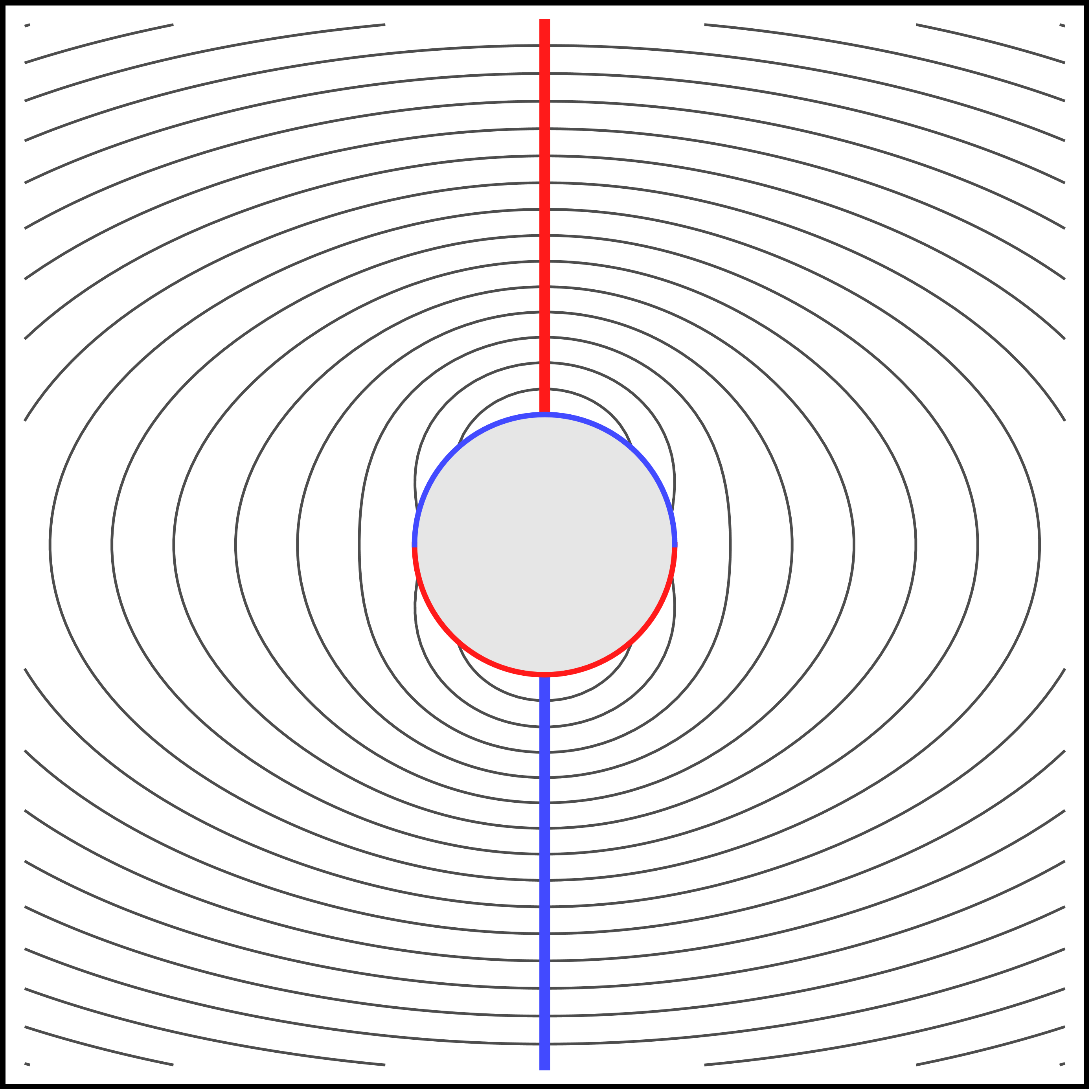}
}
\subfloat[$m=0.1,\;n=2,\;a=0.9.$]{%
    \includegraphics[width=0.23\textwidth]{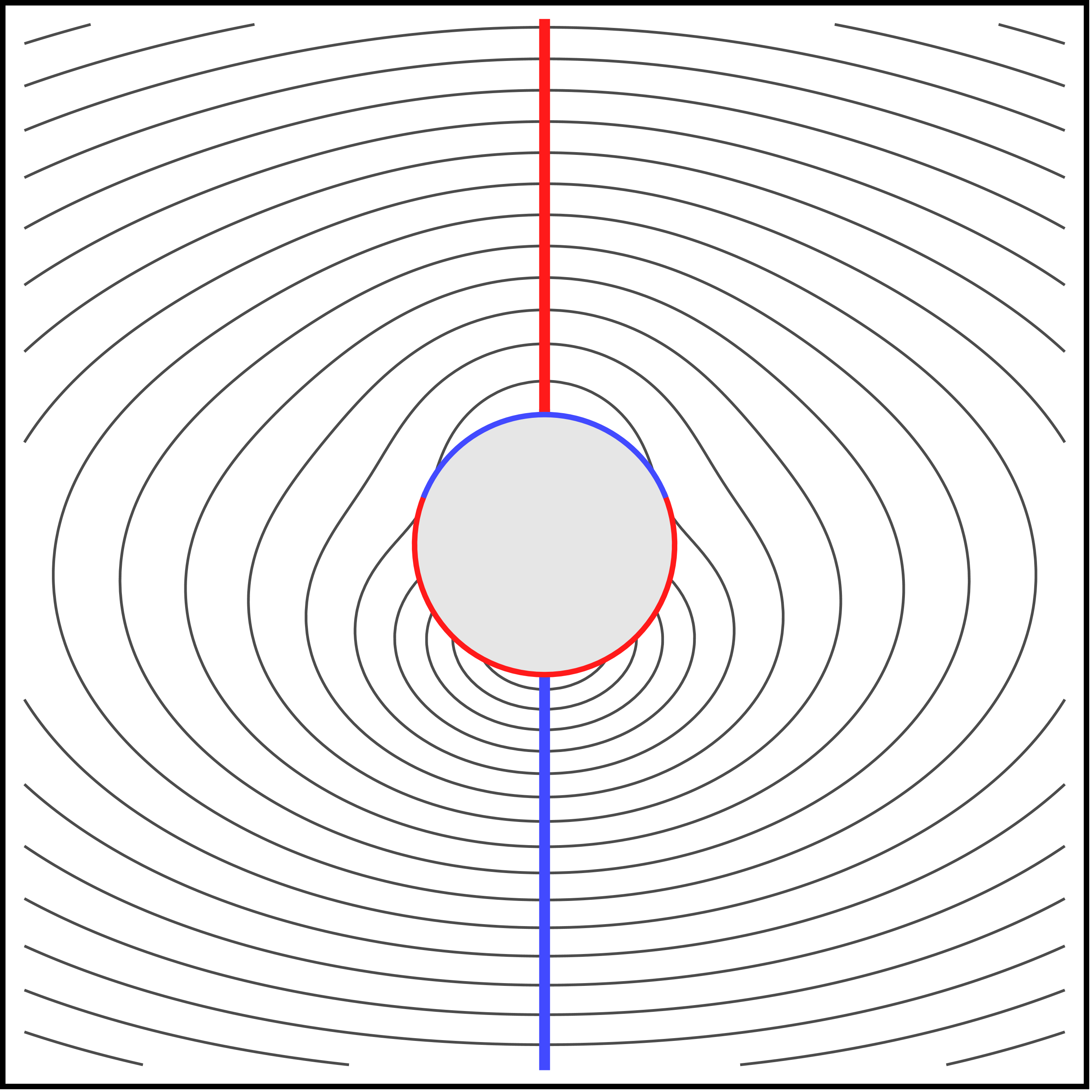}
}
\caption{Exact FL patterns of the Komar two form $d\tilde{k}$, as measured by a ZAMO observer
for (a) Taub-NUT and (b) Kerr-NUT solutions . We see that the presence of a nonzero rotation parameter $a$ leads to the asymmetry with respect to the equatorial plane. Moreover, now there are FLs escaping to infinity and crossing $S_r^\infty$. We leave more detailed patterns for a separate publication.}
\label{FIG 2}
\end{figure}

There are several approaches to studying the thermodynamics of black holes with a NUT parameter \cite{Clement:2019ghi,Bordo:static,Bordo:rotating,Liu}, which differ in formulating the Smarr relation. In \cite{Bordo:static}, the asymptotic mass plays the role of mass in the Smarr relation, while the contribution from the Misner strings is expressed in   terms of “gravitational Misner charges” $ N_\pm$ and the “Misner potentials” $\psi_\pm$. In fact, the quantities $\psi_+ N_+$ and $\psi_- N_-$ coincide  exactly with the half of the fictitious Misner string masses $M^\pm_\text{fict}$, found in \cite{Galtsov:2026wxl}. Later, this construction was generalized to the rotating case in the paper \cite{Bordo:rotating}. In \cite{Liu}, thermodynamics is investigated by applying the Wald formalism to the Taub-NUT solution. The mass $M$ appearing there in the Smarr relation and the first law coincides with our horizon mass $M_H$ from \cite{Galtsov:2026wxl}. In this approach, the Smarr relation does not contain terms associated with Misner strings. Expressions for $M^\pm_\text{fict}$ and $M_H$ were first derived in \cite{Clement:2019ghi} based on the Komar–Tomimatsu approach—the "rod structure" formalism. We leave the discussion of thermodynamics from the perspective of the current approach for a separate publication.

\section{Cartan equations}
 An important observation is that both MSs are entirely within the chronosphere, where the Killing vector $\partial_\varphi$ is timelike \cite{Kolar:2025kle}. Spacelike azimuthal Killing vectors within the chronosphere can be constructed using the rod formalism \cite{Harmark:2004rm,Clement:2019ghi}:
\begin{equation}\xi^s=\partial_\varphi-2ns\partial_t=\partial_{\varphi^s},
\end{equation}
Their norm vanishes on the strings.
Now, although the integral curves $\partial_\varphi$ are closed in the compact region $0\le\varphi\le 2\pi$, the integral curves of the space-like vector $\partial_{\varphi^s}$ are not closed. Namely, by parallel transporting this vector along the curve encompassing the polar axis, we obtain the temporary discrepancy (the Burgers vector of the dislocation):
$\Delta x^\mu=- 4\pi ns\delta^\mu_t
$ which is directly proportional to the torsion tensor and the loop area: $
\Delta x^{\mu }=- T_{\phantom{\mu }\nu \lambda }^{\mu }\Delta F^{\nu \lambda }/2$, where $\Delta F^{\nu \lambda }$ is the element of the surface, perpendicular to the dislocation axis.
Therefore, \begin{equation}
	T_{\phantom{\mu }\nu \lambda }^{\mu } = \delta^\mu_t T_{\phantom{\mu }\nu \lambda }^{t }. 
\end{equation}

Another problem of Taub-NUT is that the corresponding Levi-Civita connection is singular. To see this, we pass
to the orthonormal frame $e^a,\,a=0,1,2,3$ and write down the first Cartan equation \cite{Cai:2015emx}:
\begin{equation}\label{Cartan}
	de^a+\omega^a_{\;b}\wedge e^b=T^a, 
\end{equation}
where $T^a$ is the torsion two-form. Here $e^0=F(dt+2n\cos\theta d\varphi),\;\;F^2=\Delta/\Sigma$, so by \eqref{ddphi} the singular term is present in \eqref{Cartan}:
\begin{equation}\label{desing}(de^a)_\mathrm{ sing}=\delta^a_0 4\pi nsF\,\delta^2(\mathbf x)\,dx\wedge dy.
\end{equation}
Thus, we have two options: either take the singular Levi-Civita connection $\omega^a_{\phantom{a}b}=\omega^a_{\phantom{a}b}(e)$ with zero torsion, or choose a regular connection with non-zero torsion compensating for the singular term \eqref{desing}.
In the first case, we will get the  singular Einstein tensor :
\begin{align}
	&\frac13G^0_0\!=\!G^1_1\!=\!-G^2_2\!=\!-G^3_3=\frac{4\pi n^2\Delta}{\Sigma^3}[1-\pi\delta^{2}(\mathbf{x})]\delta^{2}(\mathbf{x}),\nonumber\\
	&G^0_2\!=\! \sigma(y\partial_x\!-\!x\partial_y) \delta^{2}(\mathbf{x}),\; G^0_3\!=\!\sigma(x\partial_x\!+\!y\partial_y) \delta^{2}(\mathbf{x}),
\end{align}
where $\sigma=-2\pi sn\sqrt{\Delta}/\sqrt{x^2+y^2}\Sigma^2$. The off-diagonal terms admit a reasonable interpretation, reproducing the rotation term $R^a_t$ on MS found in \cite{Clement:2022pjr}, but the diagonal terms containing squares of delta-functions seem problematic.

It may seem paradoxical that, despite the singularity of the Levi-Civita connection, Misner strings are geodesically transparent\footnote{We thank reviewer for drawing our attention to this.} \cite{Clement:2015cxa}. Moreover, the bulk geometry of the Taub-NUT spacetime appears unaffected by the "delta-squared" type singularity in the Einstein tensor. This situation contrasts with the case of the linear delta-singularity in the energy-momentum tensor of a cosmic string, which creates a conical spacetime structure and alters the global behavior of geodesics. It appears that, in a certain sense, the "delta-squared" singularity is weaker than the standard linear singularity. Some support for this hypothesis can be found in Colombeau theory \cite{colombeau}, which allows for operations involving the squares of the delta functions. However, the specific problem addressed here has not yet been investigated within the framework of Colombeau theory. For a more detailed discussion of these topics, see  \cite{Steinbauer:2006qi}.

 \section{Torsionful connection}
Passing to the Cartesian-like coordinates on a two-sphere $x=u\cos \varphi,\;
    		y=u\sin \varphi,$ where $
 u=\sin \theta    $, we get
\begin{align}	&e^0=F\left[dt+2nsv(\partial_xV
	dy-\partial_yVdx)\right],\\ &e^1=\frac{dr}{ F},\; e^2=\frac{\sqrt{\Sigma}(xdx+ydy)}{uv}, \; e^3=\frac{\sqrt{\Sigma}(xdy-ydx)}{u},\nonumber
\end{align}
where  
$V \!=\!\ln u, \, \,v=\sqrt{1-u^2}.
$  
The set of connection one-forms admitting the torsion two-form satisfying $T^a\!=\!\delta^a_0 T^0$, reads:
\begin{align}\label{spin connection}
	&\tensor{\omega}{^0_1}=  F' e^0,\;\; \tensor{\omega}{^1_i}=-\frac{F\Sigma'}{2\Sigma} \delta_{ij}e^j, \;\; \tensor{\omega}{^0_i}=-s\kappa\epsilon_{ij} e^j,  \nonumber\\
    &\tensor{\omega}{^2_3}=-s\kappa  e^0-\frac{v}{u\sqrt{\Sigma}}  e^3,\;\;
    \kappa= \frac{nF}{\Sigma}\;\nabla_{\vec{u}}V,
\end{align} 
where $i,j\!=\!2,3,\,\epsilon_{23}\!=\!1$ and $\nabla_{\vec{u}}V\!\!=\!x\partial_xV\!+\!y\partial_yV$. The non-zero torsion component is:
\begin{equation}
\tensor{T}{^0_{23}}=2sn\frac{v^2F }{\Sigma}(\partial^2_xV+\partial^2_yV)  
	=4\pi sn \frac{F}{\Sigma}\delta^2(\mathbf{x}).
\end{equation}
In the coordinate basis the only non-zero  component is:  
\begin{equation}\label{torsion TNUT}
	\tensor{T}{^t_{xy}}=4\pi  sn \delta^2(\mathbf{x}).
\end{equation}
Now, calculating the curvature 2-forms of the torsionful connection \eqref{spin connection}:
\begin{equation}
	\tensor {\mathcal {R}} {^a_b} = \frac {1} {2} \tensor {R} {^a_{bcd}} e^c \wedge e^d = d \tensor {\omega} {^a_b} + \tensor {\omega} {^a_c} \wedge \tensor {\omega} {^c_b}, 
	\end {equation}
	taking into account $\partial^2_xV+\partial^2_yV=2\pi\delta^2(\mathbf{x})$, and performing contraction to obtain the Einstein tensor, we find that its only nonzero frame components are given by:
	\begin{equation}\label{Greg}
		G^0_0 = G^1_1 =-4\pi\frac{\lambda_m}{\Sigma}\delta^2(\mathbf{x}), \quad \lambda_m =-\frac{n^2\Delta}{\Sigma^2},
	\end{equation}
were $\lambda_m$ coincides with a   (fictitious) mass density along  MS introduced in \cite{Galtsov:2026wxl}.

One can relate the torsion 2-form to the algebraic Bianchi identities by taking the exterior differential of the first Cartan equation and using the second equation:
\begin{equation}\label{Bianchi with torsion}
dT^a+\tensor{\omega}{^a_b}\wedge T^b=\tensor{{\cal R}}{^a_b}\wedge e^b.
\end{equation}
It was proposed \cite{Argurio:2009xr} that, to preserve gravitational duality, it is necessary to violate the algebraic Bianchi identities by introducing the energy-momentum tensor of "dual matter" $\Theta_{\mu\nu}$  as the source via the relation: 
\begin{equation}\label{dual EMT} R_{abcd}+R_{adbc}+R_{acdb}=8\pi \Theta_{am}\tensor{\epsilon}{^m_{bcd}}.
\end{equation}
Taking the Hodge dual from both sides of the Bianchi equation with torsion, we obtain:
\begin{equation} \star(dT^a+\tensor{\omega}{^a_b}\wedge T^b)=8\pi \tensor{\Theta}{^a_b}e^b.
\end{equation}
Using this, we finally obtain the nonzero components of the  dual Einstein tensor  defined via the dual Riemannian curvature   \cite{Argurio:2009xr}, entering the  dual   Einstein equation $\tilde{G}_{ab}=8\pi\Theta_{ab}$: 
\begin{equation}
\tilde{G}^0_0=\tilde{G}^1_1= -4\pi\frac{\lambda_n}{\Sigma}\delta^2(\mathbf{x}),
\end{equation}
which is similar to \eqref{Greg}, but now with  the fictitious dual mass density $\lambda_n$ \eqref{lamn} along the MS.

The invariance of the equations of motion under the $SO(2)$ duality rotation  of the curvature 2-form ${\cal R}_{ab}$ with its Hodge dual $\tilde{{\cal R}}_{ab}=1/2\epsilon_{abcd}{\cal R}^{cd}$ for theories with torsion was revealed in \cite{Kol:2023yxd}. Since we now have
	\begin{equation}
		\tilde{{\cal R}}_{ab}\wedge e^b=-G_{ab}\star e^b,\qquad {\cal R}_{ab}\wedge e^b=\tilde{G}_{ab}\star e^b,
	\end{equation}
 it is easy to see that this duality in the Taub-NUT solution reduces to the $SO(2)$  rotation of  $\lambda_m$ and $\lambda_n$.
 
 In Einstein-Cartan theory, the torsion tensor is related by the equations of motion to the spin angular momentum tensor $\tensor{\tau}{^\sigma_{\mu\nu}}$ by the algebraic relation:
\begin{equation}\label{spin dens EC}
\tensor{T}{^\sigma_{\mu\nu}}+\delta^\sigma_\mu \tensor{T}{^\lambda_{\nu\lambda}} + \delta^\sigma_\nu\tensor{T}{^\lambda_{\mu\lambda}}=8\pi \tensor{\tau}{^\sigma_{\mu\nu}}.
\end{equation}
For the torsion tensor \eqref{torsion TNUT}  the last two terms on the left-hand side vanish, so the only non-zero component is: 
\begin{equation}
\tensor{\tau}{^t_{xy}}=\frac{1}{8\pi}\tensor{T}{^t_{xy}}=s\frac{n}{2}\delta^2(\mathbf{x}).
\end{equation}

The natural way to describe rotating matter in Einstein-Cartan theory is to use the Weyssenhoff semiclassical spin-fluid model \cite{Weyssenhoff:1947vye, Obukhov:1987yu}. This fluid is a continuous medium possessing intrinsic angular momentum described by the spin density tensor $s_{\mu\nu}$, while the energy density and pressure are specified by the energy-momentum tensor of a perfect fluid. There is a simple relationship between the tensor $s_{\mu\nu}$, the fluid 4-velocity field $u^\mu$, and the spin angular momentum tensor:
\begin{equation}\tensor{\tau}{^\sigma_{\mu \nu}}=s_{\mu \nu}u^\sigma,
\end{equation}
In our case, fluid is at rest,  $u^\mu=\delta^\mu_t/F$,  
so for the spin density we obtain:
\begin{equation}\label{spin dens}
s_{xy}(r)=s\frac{n}{2}F\delta^2(\mathbf{x}),
\end{equation}
which asymptotically coincides with the Komar angular momentum density \eqref{Komar spin dens}:
\begin{equation}
s_{xy}(\infty)=\lambda_J^s(\infty)=s\frac{n}{2}.
\end{equation}

The energy-momentum tensor of the anisotropic Weyssenhoff fluid reads \cite{Obukhov:1987yu}:
\begin{equation}T^{\alpha \beta}=\varepsilon e_0^\alpha e_0^\beta+p_ie_i^\alpha e_i^\beta, 
\end{equation} 
where  $\varepsilon$ and $p_i$ stand for energy density and anisotropic pressure, we can apply the Einstein equations to extract these quantities from \eqref{Greg}. Clearly, only one component of the pressure $p=p_1$ is non-zero, $p_2=p_3=0$, and  
	\begin{equation}\label{ener dens}
		\varepsilon=-p=\frac{\lambda_m}{2\Sigma}\delta^2(\mathbf{x}).
	\end{equation}
	Thus, the source of the Taub-NUT solution is the  Weyssenhoff fluid beam, which has an equation of state similar to that of a cosmic string with variable  negative tension \eqref{ener dens} and spin density \eqref{spin dens}.  
 
\section{Recovering the Komar forms}
Now we would like to rewrite $dk$ and $d\tilde{k}$ in terms of torsion-related quantities. In the frame basis, we have:
\begin{equation}
	dk=d(k_ae^a)=dk_a\wedge e^a+k_ade^a.
\end{equation}
Introducing the covariant exterior differential:
\begin{equation}
	Dk_a=dk_a+\omega_{ab}k^b,
\end{equation}
and, using the first Cartan equation \eqref{Cartan}, we  obtain:
\begin{equation}\label{EC dk}
	dk=Dk_a\wedge e^a + k_aT^a.
\end{equation}
To ensure consistency with the previous discussion \cite{Galtsov:2026wxl}, let us evaluate this for the timelike Killing vector 1-form $ k_{(t)}\!=\!-\!Fe^0$. The first term expands as: 
\begin{equation}
	Dk_a\wedge e^a=\left(-\delta^0_am\frac{\tilde{\Delta}}{\Sigma^2}e^1+\sqrt{\frac{\Delta}{\Sigma}}\tensor{\omega}{^0_a}\right)\wedge e^a,
\end{equation}
and, further using the  spin connection, we find 
\begin{equation}
	Dk_a\wedge e^a=2m\frac{\tilde{\Delta}}{\Sigma^2}e^0\wedge e^1+2n\frac{\Delta}{\Sigma^2}e^2 \wedge e^3,
\end{equation}
which is nothing but the bulk part of the Komar 2-form $K$, presented in \cite{Galtsov:2026wxl}. The second term in \eqref{EC dk} reads:
\begin{equation}
	k_aT^a=-4\pi s n\frac{\Delta}{\Sigma}\delta^2(\mathbf{x})dx \wedge dy,
\end{equation}
which is the singular field ${\cal K}^s$ obtained in \cite{Galtsov:2026wxl}. Finally, we see that within the framework of   Einstein-Cartan theory, it is torsion that generates the MS singular terms.

\smallskip
\section{Conclusions}
 
Although the use of torsion to interpret the Taub-NUT solution has been repeatedly proposed in the literature \cite{Bonnor:2001mh,bonnor2001,Griffiths:2009dfa,Kol:2020zth,Kol:2023yxd,Kolar:2025kle}, a detailed analysis has not been conducted at this time. This became possible only after the recognition of the role of singular Komar flows in Misner strings \cite{Galtsov:2026wxl} and is summarized here.   We have shown that, within general relativity, the Levi-Civita connection of the Taub-NUT metric is singular on the Misner strings, leading to an ill-defined Einstein tensor. This singularity can be eliminated by introducing non-dynamical torsion, which compensates for the singularities in the Cartan equations. Torsion generates Komar fluxes in Misner strings, which are responsible for the asymptotic charges. The Einstein tensor thus has a structure corresponding to the stress tensor of cosmic strings with negative variable tension. The dual Einstein tensor, constructed using the dual curvature tensor, has a similar structure, demonstrating the S-duality of the solution.

Torsion naturally explains the presence of a chronology-violation region around  Misner strings, reproducing the Burgers vector arising from the translation of the spacelike azimuthal Killing vector around the polar axis. 

In the framework of the classical Einstein-Cartan theory, the sources of distribution-valued torsion in Misner strings are two stationary spin-fluid beams with the equation of state $p=-\varepsilon$. It remains an open question whether the Taub-NUT solution and similar ones that exhibit gravity duality might be better suited to supergravity, where torsion is an integral part of the theory.

\section{Acknowledgements}
The authors thank Gérard Clément for helpful comments and discussions, as well as the anonymous reviewers for valuable suggestions. The work was supported by the
Foundation for the Advancement of Theoretical Physics
and Mathematics ’BASIS’.

\end{document}